\documentclass[amsmath,amssymbs,aps,twocolumn]{revtex4-2}

\usepackage{graphicx}
\usepackage{amsmath,amsfonts,amssymb}
\usepackage{epsfig}
\usepackage{wrapfig}
\usepackage{bbm}
\usepackage[usenames]{color}
\usepackage{array}
\usepackage{times}
\usepackage{float}

\def\w{\omega}

\def\a{a}
\def\ad{a^{\dag}}

\begin{document}

\preprint{APS/123-QED}

\title{Collective Topological Dark Solitons, Platicons, and Bright-Like Solitons in Normal Dispersion Optical Frequency Combs}

\author{Seyed Danial Hashemi}
\author{Ankitkumar Maisuriya}
\author{Sunil Mittal}
\email[Email: ]{s.mittal@northeastern.edu}

\affiliation{Department of Electrical and Computer Engineering, Northeastern University, Boston, MA 02115, USA}
\affiliation{Institute for NanoSystems Innovation, Northeastern University, Boston, MA 02115, USA}

\begin{abstract}
Bright dissipative Kerr solitons generated in anomalous-dispersion microresonators underpin integrated optical frequency combs that have enabled applications including precision metrology, spectroscopy, coherent communications, and optical frequency synthesis. In contrast, Kerr resonators operating in the normal-dispersion regime can support dark solitons and platicons, but even the initiation of frequency comb generation typically requires auxiliary mechanisms, such as avoided mode crossings or pulsed pumping, to satisfy the phase-matching condition for four-wave mixing. Here we show that topological edge states in two-dimensional Kerr resonator lattices intrinsically satisfy this phase-matching condition, enabling the generation of optical frequency combs and novel coherent dissipative structures in the normal-dispersion regime without any auxiliary mechanisms. The resulting frequency combs self-organize into collective dark solitons that exhibit large-scale spatiotemporal synchronization across multiple resonators at the entire lattice edge. By tuning the pump detuning, we demonstrate the continuous evolution of topological dark solitons into topological platicons while preserving collective synchronization. Remarkably, we show that the topological lattice also supports collective bright-like soliton states despite all constituent resonators operating in the normal-dispersion regime. Our results establish topological Kerr resonator arrays as a versatile route for engineering a broad family of coherent dissipative structures, ranging from dark solitons to bright-like soliton states, within a single platform. They also provide a route toward integrated frequency-comb generation in material platforms and wavelength ranges where normal material dispersion has traditionally constrained comb generation in conventional single-resonator systems.
\end{abstract}

\maketitle

The generation of optical frequency combs in Kerr microresonators exhibits a remarkably rich landscape of coherent spatio-temporal dissipative structures, including solitons, Turing rolls, and breather states \cite{Kippenberg2011, Kippenberg2018, Gaeta2019, Del-Haye2007,Herr2012,Lucas2017,Herr2014}. The most extensively studied regime is that of bright dissipative solitons, which emerge when Kerr nonlinearity balances the anomalous group-velocity dispersion of the resonator \cite{Chembo2013, Herr2014}. These states, generated in integrated photonic resonators, have enabled applications of coherent frequency combs in precision metrology, spectroscopy, coherent communications, and ultra-low-noise optical frequency synthesis on a chip \cite{Diddams2020, Suh2016,  Marin-Palomo2017, Spencer2018, Riemensberger2020,Pfeifle2014,Jia2025}. In contrast, resonators operating in the normal-dispersion regime can support dark solitons and platicons, which correspond to localized intensity dips or flat-top pulses, respectively, on a continuous-wave background in the resonator \cite{Matsko2012, Liang2014, Huang2015, Xue2015, Godey2014, Parra-Rivas2016, Parra-Rivas2017, Kostet2021, Anderson2022, Lihachev2022}. The emergence of dark solitons and platicons with normal dispersion significantly relaxes constraints on waveguide geometries. For example, it allows the use of thin silicon-nitride films to generate combs, which have been shown to support ultra-low propagation losses \cite{Bose2024, Lihachev2022}. Moreover, it enables the generation of combs at visible wavelengths, where normal material dispersion often dominates over waveguide dispersion. However, the generation of dark solitons and platicons is challenging because it typically requires additional mechanisms, such as avoided mode-crossings between different transverse modes of the resonator, pulsed driving, or dispersion engineering through photonic-crystal or photonic molecule structures to even initiate comb generation \cite{Liu2014, Jang2016, Kim2017, Kim2019, Yuan2023,  Liu2025, Ji2026, Song2026,  Yu2022, Li2023, Lobanov2025}.


\begin{figure*}[ht!]
 \centering
 \includegraphics[width=0.9\textwidth]{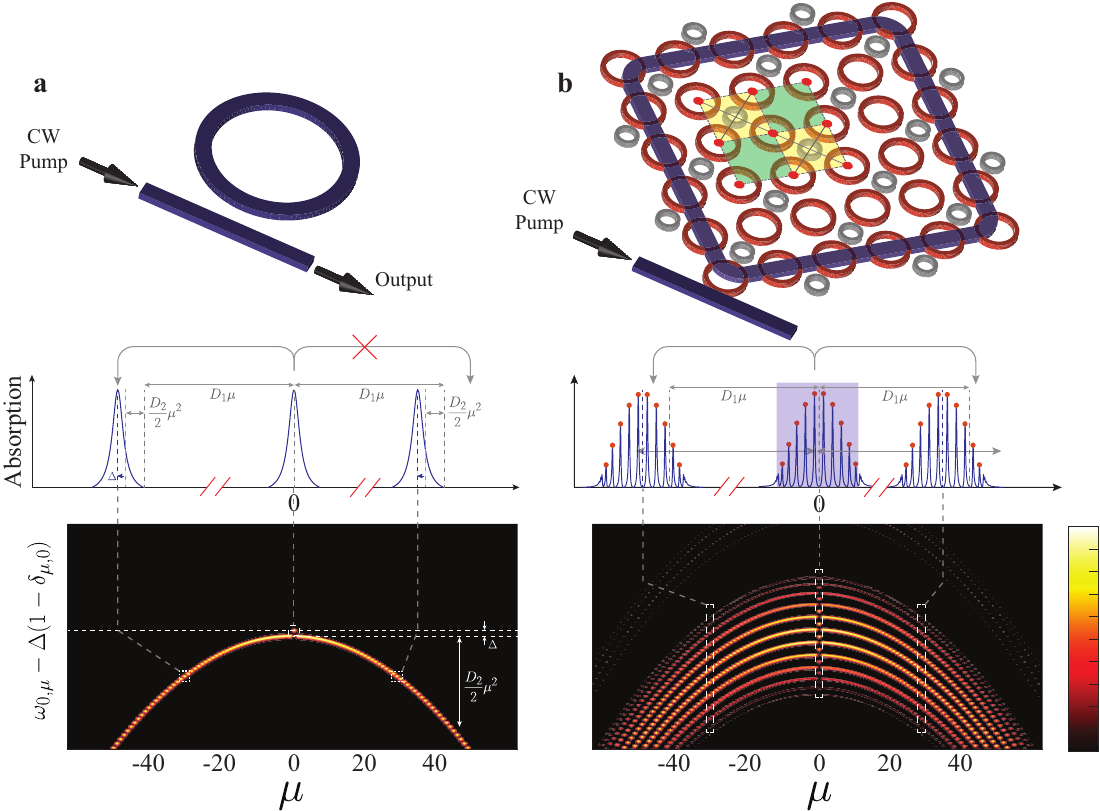}
 \caption{\textbf{Generation of optical frequency combs in coupled resonator arrays with normal dispersion.} \textbf{a.} Schematic of a single-ring resonator, with its absorption spectrum and dispersion. Due to normal dispersion and the cross-phase modulation-induced resonance frequency shift $\left(\Delta\right)$, the phase-matching condition for four-wave mixing (FWM) is not satisfied for any FSR $\mu$, preventing frequency comb generation. \textbf{b.} Schematic of the topological ring resonator lattice, with its absorption spectrum and dispersion. The presence of multiple topological edge-state resonances within each FSR intrinsically satisfies the phase-matching condition, enabling frequency comb generation without auxiliary mechanisms for initiation.}
 \label{fig:1}
\end{figure*}

Large arrays of coupled resonators have recently introduced a new paradigm for generating unconventional optical frequency combs and coherent spatio-temporal dissipative structures that are phase-locked across 10s of resonators \cite{Mittal2021b, Tusnin2023, Vasco2019, Flower2024, Xu2025, Hashemi2024, Hashemi2025, Mehrabad2025, Tikan2022}. In particular, topological edge states in two-dimensional (2D) arrays can generate nested combs and nested solitons, a comb-within-a-comb and pulse-within-pulse structure that operates at two vastly different time and frequency scales and is spatially confined to the edge of the array \cite{Mittal2021b, Flower2024, Xu2025}. Floquet-engineered topological arrays, with strong coupling between the resonators, can similarly generate incommensurate combs, in which sets of uniformly spaced comb lines are separated by an incommensurate gap \cite{Hashemi2024}. These combs are associated with the emergence of phase-locked soliton molecules spatially confined to the edge of the array. Beyond dispersion engineering, non-Hermitian dissipation engineering through topological windings has also been used to reconfigure the spectra of multi-resonator frequency combs \cite{Hashemi2025, Wang2021}. Nevertheless, these multi-resonator comb platforms have thus far relied on constituent resonators operating in the anomalous-dispersion regime. The feasibility of generating coherent spatio-temporal structures in arrays with normal-dispersion resonators has remained unexplored.

Here, we demonstrate the emergence of collective topological dark solitons, platicons, and bright-like soliton states in the edge states of a two-dimensional (2D) Kerr resonator array whose constituent resonators operate entirely in the normal-dispersion regime. We show that the presence of multiple topological edge-state resonances within each free-spectral range (FSR) provides an intrinsic phase-matching mechanism that enables frequency-comb generation and the subsequent formation of coherent dissipative structures without auxiliary mechanisms that are typically required to initiate comb generation in normal-dispersion single-resonator systems. The topological dark solitons and platicons exhibit localized intensity dips and flat-top pulses on a continuous-wave background, respectively, similar to their conventional single-resonator counterparts. However, unlike conventional dark solitons and platicons, these states are overlaid with a periodic modulation arising from the excitation of sideband edge-state resonances. More importantly, they form collective spatiotemporal states that exhibit large-scale self-synchronization and phase locking across all resonators along the lattice edge. Remarkably, even though every constituent resonator operates in the normal-dispersion regime, we show that the same topological lattice also supports collective bright-like soliton states, thereby enabling dissipative structures spanning dark solitons, platicons, and bright-like solitons within a single platform.

Although the spatio-temporal synchronization of topological dark solitons and platicons at the lattice edge is similar to that observed in topological bright nested solitons \cite{Mittal2021b}, their underlying frequency comb and spatio-temporal structure is fundamentally different. The comb spectrum associated with topological dark solitons is predominantly single-moded within each free-spectral range. In contrast,  topological bright nested solitons exhibit multiple oscillating modes within every FSR, giving rise to their characteristic comb-within-a-comb structure. Similarly, in the time domain, topological dark solitons and platicons consist of recurring single intensity dips on an oscillating background, whereas bright nested solitons exhibit a hierarchy of pulses with the characteristic pulse-within-a-pulse structure. Therefore, our results establish nonlinear topological 2D resonator arrays as a versatile platform for engineering unconventional coherent dissipative structures across both normal- and anomalous-dispersion regimes. They open new opportunities for nonlinear photonics and could enable broadband frequency comb generation at visible wavelengths, where normal material dispersion typically precludes the generation of conventional bright solitons.

\begin{figure*}[ht!]
 \centering
 \includegraphics[width=0.9\textwidth]{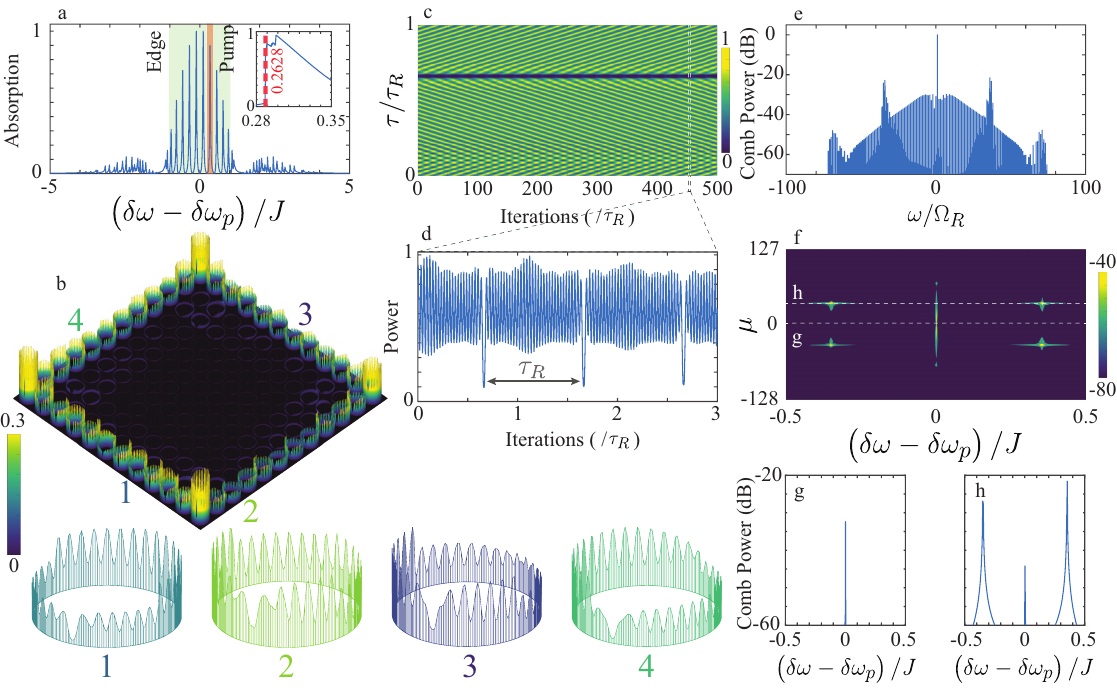}
 \caption {\textbf{Topological dark solitons.} \textbf{a.} Linear absorption spectrum of the topological lattice with edge state resonances shaded green. The pumped edge mode is shaded red. The inset shows the nonlinear cavity power. \textbf{b.} Intensity distribution in the lattice showing formation of topological dark solitons in every resonator on the edge. The insets show zoomed-in views at four different resonators. \textbf{c,d.} Temporal output showing dark soliton intensity dips, repeating every $\tau_{R}$. \textbf{e-h.} Comb spectrum showing dominant oscillation of a single edge mode at $\mu = 0$, and two side-band edge modes at $\mu = \pm 35$.}
 \label{fig:2}
\end{figure*}

Our system comprises a two-dimensional lattice of ring resonators that implements a Haldane-like anomalous quantum Hall model for photons \cite{Mittal2019, Leykam2018, Haldane1988}. The site rings, which sit at the sites of a square lattice, are coupled to their nearest and/or next-nearest neighbors via link rings (Fig.\ref{fig:1}b). The resonance frequencies of the link rings are detuned from those of the site rings such that the link rings introduce a synthetic magnetic field, with flux $\phi = \pm \pi/4$, when the photons hop between nearest-neighbor site rings \cite{Hafezi2011, Hafezi2013}. The link rings are arranged such that the flux of the synthetic magnetic field in a single plaquette is $\pi$ but that in a unit cell of the lattice is zero, similar to that in the Haldane model \cite{Mittal2019, Leykam2018}. In the limit of weak coupling between the rings and no Kerr nonlinearity, this configuration can be described by the tight-binding Hamiltonian

\begin{align}
H_{L} ={}&
\sum_{m,\mu} \w_{0,\mu} \ad_{m,\mu} \a_{m,\mu} - J \sum_{\langle m,n\rangle,\mu} \ad_{m,\mu} \a_{n,\mu} e^{-i\phi_{m,n}}  \nonumber\\
& - J \sum_{\langle\!\langle m,n\rangle\!\rangle,\mu} \ad_{m,\mu} \a_{n,\mu} + \mathrm{h.c.}.
\label{Eq_HL}
\end{align}

This Hamiltonian generates multiple copies of the anomalous quantum Hall model, one at each longitudinal mode resonance of the constituent rings. The longitudinal modes are labeled by the index $\mu$, such that $\mu = 0$ labels the pumped mode. The resonance frequencies of the corresponding modes are $\omega_{0,\mu}$. $a_{m,\mu}$ is the field in the ring at site $m$ and mode index $\mu$. $J$ is the effective coupling strength between the nearest and next-nearest site rings. The absorption spectrum of this lattice, through an input-output waveguide coupled to one of its corners, exhibits a topological edge band occupied by multiple edge state resonances. Each edge state resonance corresponds to a given longitudinal mode of the edge-state super-resonator formed at the boundary of the lattice. The neighboring sets of edge band resonances are separated by the FSR $D_{1} = \Omega_{R}$ of the constituent single-ring resonators.

To provide an intuitive picture for the emergence of frequency combs in topological lattices with constituent resonators operating in the normal-dispersion regime, we compare the interplay between chromatic dispersion and the pump-induced self- and cross-phase-modulation (SPM/XPM) resonance shifts in a single resonator and in a topological lattice. For single-pump four-wave mixing, the resonance frequency of the pumped mode is shifted by the SPM contribution, proportional to $\left|a_{0}\right|^{2}$, whereas the sideband resonances experience an XPM shift proportional to $ 2\left|a_{0}\right|^{2}$ (Fig.\ref{fig:1}a). In a single resonator, the combination of these differential nonlinear resonance frequency shifts and normal dispersion detunes sideband resonances away from the four-wave-mixing phase-matching condition (indicated by the dashed straight line) and, therefore, prevents the initiation of parametric oscillation. In contrast, a topological lattice supports multiple edge-state resonances within each free-spectral range. As long as the total edge-band width, determined by the inter-resonator coupling $J$, exceeds the dispersion-induced detuning $D_{2}/2 \mu^{2}$ at a given mode number, one or more edge-state resonances satisfy the four-wave-mixing phase-matching condition (Fig.\ref{fig:1}b). Therefore, the presence of multiple edge states in the topological edge band provides an intrinsic mechanism for phase-matching. As a result, frequency comb generation can be initiated without auxiliary techniques, such as avoided mode crossings or pulsed pumping, that are typically required in normal-dispersion single-ring resonators.

While the intrinsic phase-matching mechanism of the edge band enables the onset of parametric oscillation and frequency comb generation, it does not, by itself, guarantee the formation of coherent dissipative structures, such as dark solitons, platicons, or bright-like solitons, that additionally require phase locking among the generated comb lines. Moreover, topological counterparts of dark solitons, platicons, and bright-like solitons constitute collective states that require nonlinear self-organization and phase locking across multiple resonators along the lattice edge. In the following, we show that the topological edge states not only enable frequency comb generation but also provide mechanisms for self-organization into collective soliton states that are phase locked across multiple rings on the lattice edge.


\begin{figure*}[ht!]
 \includegraphics[width=0.9\textwidth]{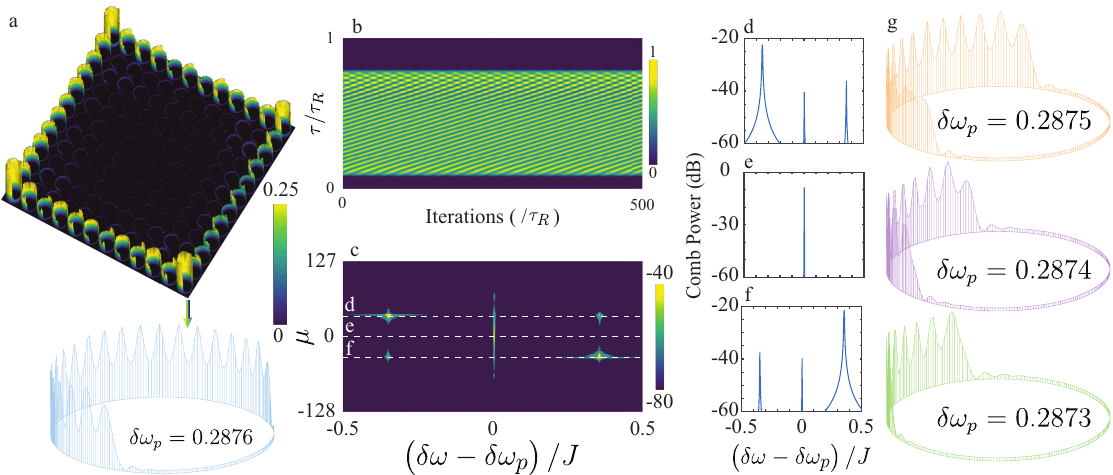}
 \caption{\textbf{Topological platicons} \textbf{a.} Lattice intensity distribution showing the formation of topological platicons in every edge resonator. The inset shows intensity in the resonator coupled to the input-output waveguide. \textbf{b,c.} Temporal and spectral output. \textbf{d,e,f.} Zoomed in comb spectra at $\mu = +35, ~0, ~-35$, respectively. \textbf{g.} Intensity distribution in the input-output ring showing tunability of the platicon width by red detuning the pump frequency. }
 \label{fig:3}
\end{figure*}

To demonstrate the formation of collective topological soliton states, we consider a $12 \times 12$ topological lattice. Its constituent resonators operate in the normal-dispersion regime with the dispersion parameter $D_{2} = 1 \times 10^{-5}$, normalized to the FSR, $v_{g}/L$, where $V_{g}$ and $L$ are the group velocity and the length of the resonator, respectively. The resonance frequencies $\omega_{0,\mu}$ of the rings are described as
\begin{equation}
\omega_{0,\mu} = \omega_{0,0} + D_{1} \mu + \frac{D_{2}}{2} \mu^{2},
\end{equation}
with $D_{1} = \left(2\pi \right) v_{g}/L$. 
We consider the intrinsic loss rate (normalized) of the resonators to be $\kappa_{in} = 3\times10^{-4}$, the coupling strength between the rings $J = 0.03$, and that between the lattice and input-output waveguide $\kappa_{ex} = 0.003$. These parameters are typical for, for example, silicon-nitride resonators where $D_{1} = \left(2 \pi\right) 250 ~\text{GHz}$, $D_{2} \sim \left(2 \pi\right) 2.5 ~\text{MHz}$, $\kappa_{in} \sim \left(2 \pi\right) 75 ~\text{MHz}$, $\kappa_{ex} \sim \left(2 \pi\right) 750 ~\text{MHz}$, and $J \sim \left(2 \pi\right) 7.5 ~\text{GHz}$ \cite{Flower2024}. A continuous-wave pump is injected into the lattice via an input-output waveguide coupled to a single resonator at one of its corners. The generated frequency comb is measured at the output of the same input-output waveguide. We use the Ikeda map formalism to simulate the generation of the comb in the lattice \cite{Hansson2016, Ikeda1979, Hashemi2024, Hashemi2025} (see Supplementary Information). We tune the pump power and pump frequency detuning to reveal the self-organization of the generated comb into collective dark soliton, platicon, or bright-like soliton regimes.

We first demonstrate the formation of topological dark soliton states. We tune the pump frequency $\delta\omega_{p} = \omega_{p} - \omega_{0,0} \simeq 0.2885~J$ close to one of the edge state resonances (shaded red in Fig.\ref{fig:2}a), and the normalized pump amplitude $E_{\text{in}} = 0.0469$ (see Supplementary). In the lattice, the spatio-temporal intensity distribution reveals a collective dark soliton state where each resonator on the edge hosts a dark soliton: a localized intensity dip on an otherwise bright background (Fig.\ref{fig:2}b) created by two locked switching wavefronts \cite{Parra-Rivas2016}. However, unlike conventional single-resonator dark solitons, the bright background of the dark soliton state exhibits pronounced periodic modulation. Remarkably, the dark soliton dips occur at identical temporal positions in every edge resonator (except for the modulation pattern). This demonstrates the formation of a collective, self-synchronized topological dark soliton state that is phase locked across all 44 resonators on the lattice edge. The temporal evolution of this collective dark soliton state shows a redistribution of intensity in each ring on the edge, without affecting its synchronization (Movie M1). We note that pumping an edge-state resonance results in negligible excitation of bulk modes because of their weak spatial overlap and phase mismatch. Consequently, the optical field remains localized along the lattice edge \cite{Mittal2021b, Flower2024, Mittal2018}.

At the output of the lattice, the collective spatio-temporal state appears as a periodic sequence of dark pulses repeating every round-trip time $\tau_{R} = L/v_{g}$, superimposed on a periodically modulated background (Fig.\ref{fig:2}c,d). Reorganizing the temporal output in terms of fast time $\tau/\tau_{R} = [0, 1]$ and slow time (round-trip number) confirms that the dark soliton dips repeat every $\tau_{R}$ and remain coherent over successive round trips. Interestingly, the background modulation is not stationary over successive round trips, as evidenced by the characteristic slanted interference pattern in Fig.\ref{fig:2}c. In fact, the modulation waves propagate in opposite directions at the two switching fronts of the dark soliton (Movie M2), giving rise to interference fringes with opposite slopes on either side of the low-intensity region. 

In the frequency domain, the output comb spectrum exhibits a smooth envelope but with prominent horn-like spectral features around $\mu \sim \pm 37$. Reorganizing this frequency spectrum into fast (FSR index $\mu$) and slow (characterized by coupling strength $J$) frequency scales  (Fig.\ref{fig:2}f) reveals that the dominant comb envelope and the horn-like features near $\mu \sim \pm 37$ originate from the oscillation of different edge state resonances. In particular, the comb spectrum is dominated by the oscillation of a single edge state supermode at each FSR (Fig.\ref{fig:2}g), corresponding to the pumped edge state. This single-edge-mode oscillation is consistent with the observation of dark solitons at each resonator on the edge of the lattice.

However, near $\mu = \pm 37$, the supermode spectrum shows oscillation of two sideband edge state resonances (at $\sim \pm 0.35 J$). It is these oscillating edge states that lead to periodic temporal modulation of the bright background of dark soliton states. The modulation period of $0.027 \tau_{R}$ in the fast time scale corresponds to the inverse of the spectral separation ($\mu \sim 37$) of horn-like features in the fast frequency scale, establishing a direct correspondence between the temporal modulation and the excitation of sideband edge-state resonances. Furthermore, the four excited sideband edge states, located at $\delta\omega - \delta\omega_{p} \simeq \pm 0.35 J ~\text{and} ~\mu \simeq \pm 37$, form two pairs that are symmetrically placed around the pump frequency. Each pair contributes to the modulation wave at one of the two switching wavefronts, and the modulation waves generated by the two pairs propagate in opposite directions (see Fig.\ref{fig:2}c and Supplementary Information). As in conventional normal-dispersion Kerr resonators, these switching wavefronts and their modulation waves lock together to form the topological dark soliton. 

We note that the horn-like features observed in the comb spectrum are also a characteristic signature of conventional normal-dispersion dark solitons generated in single resonators, where they originate from the oscillatory tails of the two switching waves that bind together to form a dark soliton \cite{Liang2014, Parra-Rivas2017, Yu2022}. In contrast, the horn-like features of topological dark solitons originate from the excitation of neighboring topological edge-state resonances. The higher intensity of these sideband edge-state resonances gives rise to the periodic modulation of the high-intensity background surrounding the dark soliton, a feature absent in conventional single-resonator dark solitons. Thus, while both conventional and topological dark solitons exhibit horn-like spectral features, their physical origin is fundamentally different. In single resonators, they reflect the oscillatory tails of bound switching waves, whereas in topological lattices they arise from the excitation and interference of neighboring edge-state resonances.

The observed topological dark soliton states are also fundamentally different from the topological bright nested soliton that can be observed in this same lattice, but with its constituent rings operating in the anomalous dispersion regime \cite{Mittal2021b}. In the frequency domain, nested solitons exhibit the simultaneous oscillation of multiple edge-state supermodes within every FSR, giving rise to the characteristic comb-within-a-comb spectrum. In contrast, the dark-soliton state is dominated by a single edge-state supermode at each FSR. In the time domain, the temporal output of nested solitons exhibits the pulse-within-a-pulse structure, whereas that of dark solitons features synchronized intensity dips, resembling those of single-ring resonators but with a modulated background. 

\begin{figure*}[ht!]
 \centering
 \includegraphics[width=0.9\textwidth]{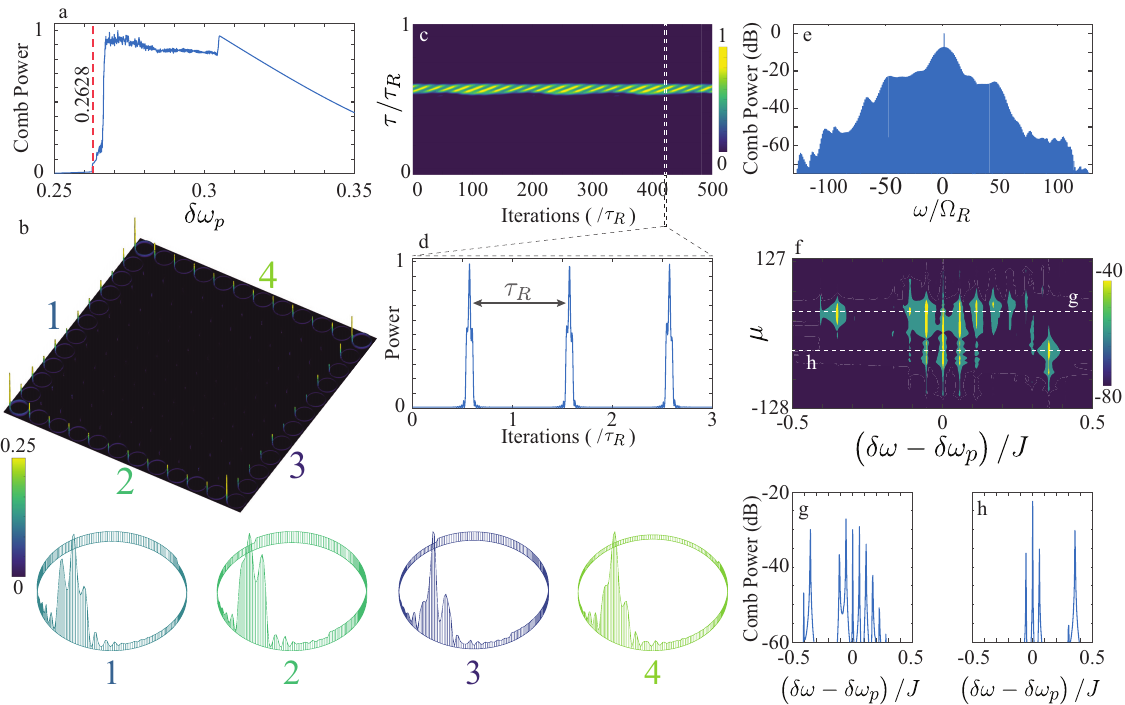}
 \caption{\textbf{Topological bright-like solitons} \textbf{a.} Nonlinear pump power as a function of pump frequency detuning. \textbf{b.} Lattice intensity distribution showing formation of topological bright-like solitons. Insets show intensity profiles in four different resonators on the edge. \textbf{c, d.} Temporal output as a function of round trips showing generation of bright pulses. \textbf{e-h.} Comb spectrum showing oscillation of multiple edge state resonances at each FSR.}
 \label{fig:4}
\end{figure*}


In normal-dispersion Kerr resonators, dark solitons have been shown to evolve continuously into platicons as the pump detuning is varied \cite{Parra-Rivas2016, Yu2022}. This evolution is accompanied by a continuous increase in the width of the low-intensity region bounded by two switching fronts, giving rise to the characteristic flat-top (or equivalently, flat-bottom) temporal profile of platicons. We observe an analogous transition in topological lattices, where collective topological dark solitons continuously evolve into collective topological platicons.

Figure \ref{fig:3} shows the formation of a collective topological platicon state, in which each edge resonator supports a synchronized platicon characterized by a broad flat-top intensity profile with a weak periodic modulation. As in the dark-soliton regime, the switching fronts remain synchronized across all edge resonators, demonstrating large-scale collective synchronization (Fig.\ref{fig:3}a and Movie M3). The temporal output therefore consists of a periodic sequence of modulated flat-top pulses repeating every round-trip time $\tau_{R}$ (Fig.\ref{fig:3}b). Nevertheless, unlike the dark-soliton state, where the modulating wave reverses its propagation direction across the two switching fronts, the modulation associated with the platicon propagates in the same direction across both fronts (Movie M4). Consequently, the interference pattern changes from oppositely slanted fringes (Fig. \ref{fig:2}c) to a single slant direction (Fig. \ref{fig:3}b).

The corresponding comb spectrum continues to be dominated by a single edge-state supermode oscillating at each free-spectral range, while weak excitation of neighboring edge-state resonances near $ \mu \sim \pm 37$ gives rise to the periodic modulation of the flat-top spatio-temporal profile (Fig.\ref{fig:3}b). In contrast to the topological dark-soliton state, however, only a single pair of sideband edge-state resonances is excited, located at $\delta\omega - \delta\omega_{p} \simeq - 0.35 J ~\text{and} ~\mu \simeq +37$ and $\delta\omega - \delta\omega_{p} \simeq + 0.35 J ~\text{and} ~\mu \simeq -37$ (Fig.\ref{fig:3}d-f, Fig.\ref{fig:2}h). This observation is in agreement with the spatio-temporal dynamics where the modulating wave propagates in a single direction at both the switching wavefronts of the platicon (Fig.\ref{fig:3}b).

Interestingly, we find that by red-detuning the pump frequency, we can continuously reduce the width of the high-intensity plateau of the platicon state. The spatio-temporal intensity profile of such platicons in the input-output resonator is shown in Fig.\ref{fig:3}g for different pump frequencies. Throughout this continuous evolution, the platicon states remain phase locked across all edge resonators, while their comb spectra and temporal dynamics remain qualitatively identical to those shown in Fig.\ref{fig:3}b,c. This behavior extends the continuous platicon dynamics observed in single resonators \cite{Parra-Rivas2016, Parra-Rivas2017, Yu2022} to collective topological states that are synchronized across multiple resonators.

While dark solitons and platicons are the hallmark coherent dissipative structures of normal-dispersion Kerr resonators, bright solitons are conventionally associated with anomalous dispersion. Remarkably, despite all constituent resonators operating in the normal-dispersion regime, the topological lattice also supports the formation of collective bright-like soliton states. To access such bright-like soliton states, we further red-detune the pump frequency to $\delta\omega_{p} = 0.2628 ~J$, while continuing to pump the same edge state resonance, and marginally increase the normalized pump amplitude $E_{in} = 0.062$. As shown in Fig.\ref{fig:4}, in the lattice, each resonator on the edge now hosts a narrow bright-like soliton pulse whose intensity is periodically modulated (insets of Fig.\ref{fig:4}b and Movie M5). Similar to the dark-soliton and platicon states, these pulses remain phase locked across all edge resonators, demonstrating large-scale collective synchronization (Fig.\ref{fig:4}b and Movie M5). The temporal output therefore consists of a periodic train of bright pulses, repeating every round-trip time $\tau_{R}$. This is in contrast to bright nested solitons with anomalous dispersion, where the temporal output exhibits pulse-in-a-pulse structure \cite{Mittal2021b}. Furthermore, unlike the dark-soliton state, the corresponding frequency comb is no longer dominated by a single edge-state supermode. Instead, multiple edge-state supermodes oscillate within each free-spectral range, producing a spectral structure that closely resembles that of previously demonstrated topological nested bright solitons. The oscillation of edge state resonances at $\delta\omega - \delta\omega_{p} \simeq \pm 0.35~J$, near $\mu \sim \pm 37$ persists and gives rise to the periodic modulation observed in the temporal intensity profile of the bright-like soliton state. We emphasize that bright soliton states can also be observed in conventional single-resonator normal dispersion combs, but only in the presence of higher-order dispersion \cite{Parra-Rivas2017, Yu2022}. Here, we do not include any higher-order dispersion for the constituent rings of the lattice. Nevertheless, the intrinsic multimode nature of the edge states enables the formation of collective bright-like solitons in this lattice.


In summary, we have demonstrated the emergence of collective dark solitons, platicons, and bright-like soliton states in topological Kerr resonator arrays whose constituent resonators operate entirely in the normal-dispersion regime. We showed that the presence of multiple topological edge-state resonances within each free-spectral range provides an intrinsic phase-matching mechanism that enables frequency-comb generation without the auxiliary techniques typically required in conventional normal-dispersion Kerr resonators. Beyond enabling parametric oscillation, the generated combs self-organize into coherent dissipative structures that are synchronized and phase-locked across the entire lattice edge, establishing a new class of collective nonlinear states distributed over many resonators. More broadly, our results demonstrate that topological resonator arrays provide a versatile platform for engineering coherent dissipative structures across both normal and anomalous-dispersion regimes. This capability opens new opportunities for exploring collective nonlinear dynamics in synthetic photonic matter, while providing a practical route toward integrated frequency-comb generation. It offers a viable path in material platforms and wavelength ranges, such as thin-film silicon nitride and the visible spectrum, where normal material dispersion has traditionally hindered conventional bright-soliton operation.

\vspace{12pt}
\noindent
\textbf{Acknowledgements}
This research was supported by NSF CAREER Grant No. 2542534 and NSF Grant No. DMR-2323908. \\

\noindent
\textbf{Author Contributions:} S.M. conceived the idea and developed the simulation framework. S.D.H. performed the numerical simulations with assistance from A.M. All authors contributed to analyzing the data. S.M. wrote the manuscript with inputs from S.D.H.. S.M. supervised the project.\\


\providecommand{\noopsort}[1]{}\providecommand{\singleletter}[1]{#1}

\end{document}


\preprint{APS/123-QED}

\title{Supplementary Information: Collective Topological Dark Solitons, Platicons, and Bright-Like Solitons in Normal Dispersion Optical Frequency Combs}

\author{Seyed Danial Hashemi}
\author{Ankitkumar Maisuriya}
\author{Sunil Mittal}
\email[Email: ]{s.mittal@northeastern.edu}

\affiliation{Department of Electrical and Computer Engineering, Northeastern University, Boston, MA 02115, USA}
\affiliation{Institute for NanoSystems Innovation, Northeastern University, Boston, MA 02115, USA}

\maketitle

\section{Modulation Wave}
\begin{figure*}
 \centering
 \includegraphics[width=0.9\textwidth]{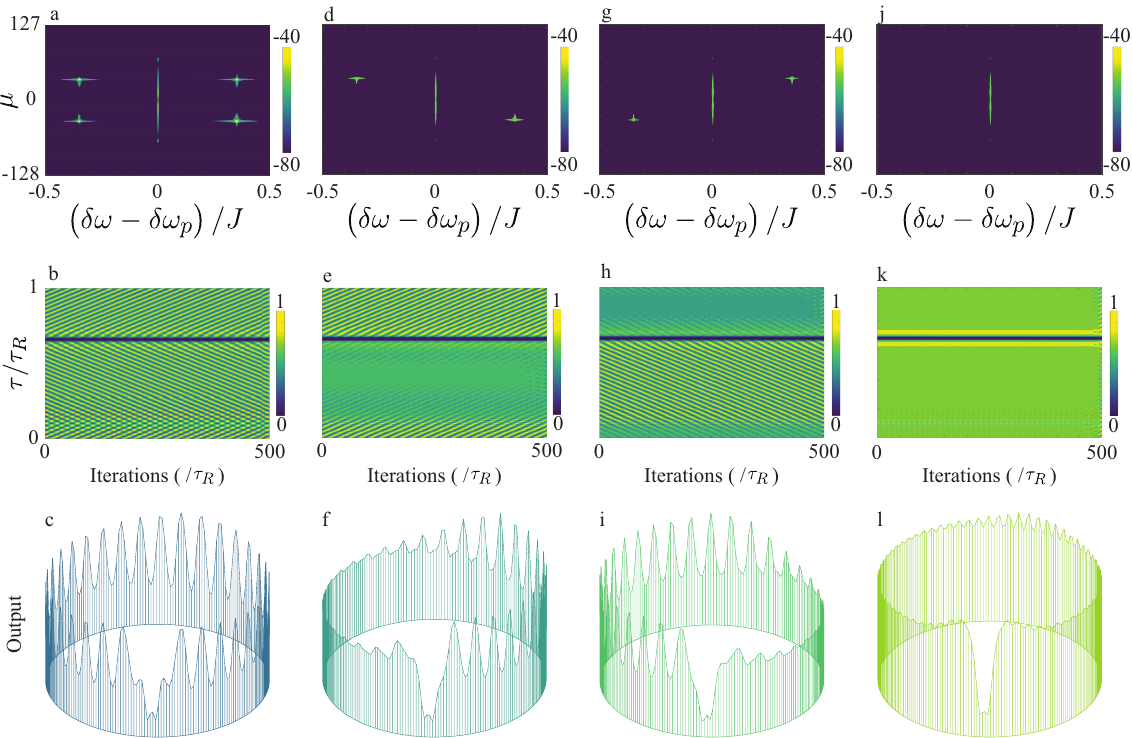}
 \caption{\textbf{Modulation of the high-intensity background of the topological dark solitons}.
\textbf{a}. Comb spectrum as a function of the fast frequency (mode index $\mu$) and the slow
frequency $(\delta\omega-\delta\omega_p)/J$, and \textbf{b,c.} corresponding spatio-temporal intensity in the input-output ring showing modulation wave at the two switching wavefronts.
\textbf{d-f} Corresponding results when we suppress one pair of oscillating sideband edge state resonances. The modulation wave at one switching wavefront is suppressed. 
\textbf{g-i} Corresponding results when we suppress the other pair of oscillating sideband edge state resonances, which suppresses the modulation wave at the other switching wavefront.
\textbf{j-l} Comb spectrum and spatio-temporal intensity distribution with both pairs of sideband edge state resonances suppressed. In this case the modulation at both wavefronts is suppressed.
}
 \label{fig:S1}
\end{figure*}

The dark solitons, platicons, and bright-like soliton states presented in the main text exhibit a periodic modulation of their intensity profiles. The modulation period, $\sim 0.027\tau_R$, is consistent with the excitation of sideband edge-state resonances near $\mu \sim \pm 37$ and $\delta\omega-\delta\omega_p \sim 0.35J$. For the dark-soliton state, two pairs of sideband resonances are excited, giving rise to modulation waves that propagate in opposite directions at the two switching wavefronts (Fig.~\ref{fig:S1}a--c). In contrast, for the platicon state, only a single pair of sideband resonances is excited, consistent with the modulation wave propagating in a single direction.

To further elucidate this behavior, we focus on the dark-soliton state, for which two pairs of sideband edge-state resonances are present. We selectively suppress each pair in the comb spectrum (Fig.~\ref{fig:S1}a,d,g,j) and reconstruct the corresponding spatiotemporal field using an inverse Fourier transform. The resulting profiles show that each symmetrically positioned pair of sideband resonances contributes predominantly to the modulation at one of the two switching wavefronts, producing a single set of diagonal fringes in the spatiotemporal intensity distribution. When both pairs of sideband resonances are suppressed (Fig.~\ref{fig:S1}j--l), the modulation vanishes from both switching wavefronts, leaving the underlying dark-soliton profile.

\section{Chaotic Comb}
\begin{figure*}
 \centering
 \includegraphics[width=0.9\textwidth]{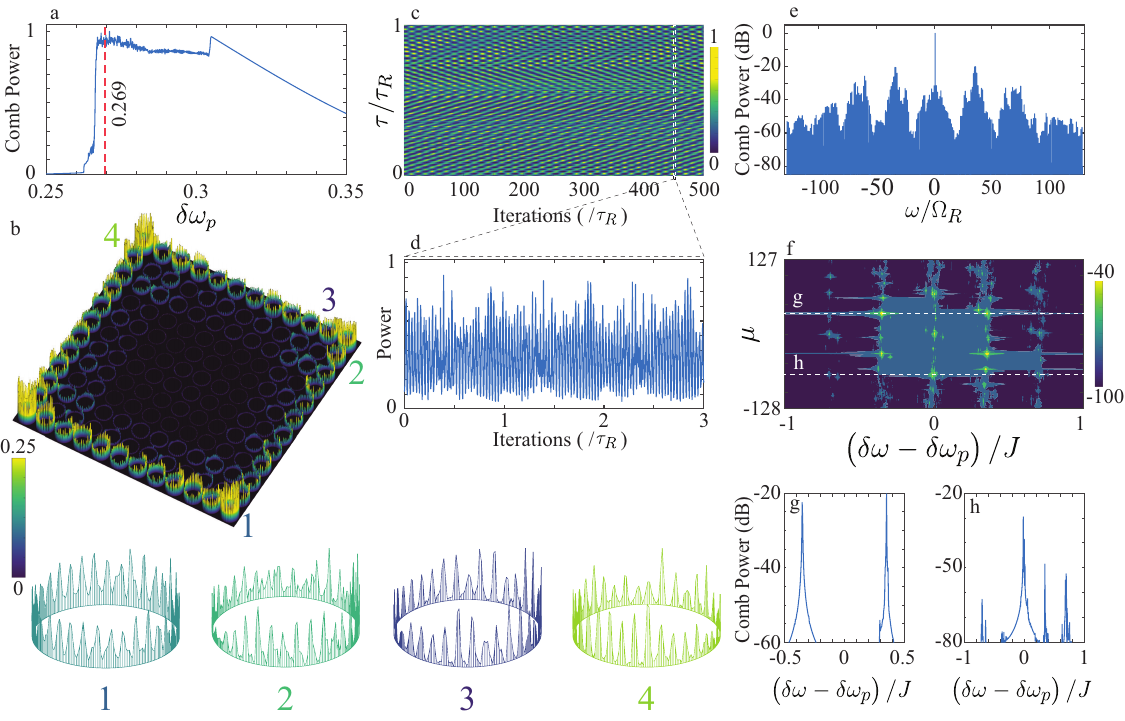}
 \caption{\textbf{Chaotic comb}
\textbf{a.} Generate comb power as a function of pump frequency detuning. The dashed red line highlights the pump frequency for the representative chaotic comb spectrum. 
\textbf{b.} Lattice intensity distribution showing absence of any spatio-temporal synchronization in a chaotic comb. Insets show intensity profiles in four different resonators on the edge. 
\textbf{c, d.} Temporal output as a function of round trips showing the lack of any periodic patterns.
\textbf{(e-h)} Comb spectrum showing noisy envelope and the oscillation of different edge states at different FSRs.}
 \label{fig:S2}
\end{figure*}

The results presented in the main text demonstrate the formation of coherent dissipative structures for specific pump-frequency detunings. However, as in single-ring Kerr resonators, other pump detunings in the topological lattice can give rise to incoherent modulation-instability or chaotic comb states, even when the same edge-state resonance is pumped. A representative chaotic comb, obtained by blue-detuning the pump relative to the operating point shown in the inset of Fig.~4 of the main text, is presented in Fig.~\ref{fig:S2}. In this regime, the spatiotemporal intensity distributions in different edge resonators no longer exhibit the collective synchronization observed for the coherent soliton states (Fig.~\ref{fig:S2}b). Nevertheless, because an edge-state resonance is pumped, the optical field remains predominantly confined to the lattice edge, with negligible intensity in the bulk. Correspondingly, the temporal output fluctuates irregularly from one round trip to the next, without a stable periodic waveform (Fig.~\ref{fig:S2}c,d).

In the frequency domain, the generated comb exhibits a broad, noisy spectral envelope (Fig.~\ref{fig:S2}e--h). Different edge states oscillate at different FSRs. However, unlike in the coherent dark-soliton and platicon states, these resonances oscillate incoherently and do not self-organize into a phase-locked collective state.


\section{Numerical Simulation}
We simulate the full nonlinear frequency-comb dynamics of the topological resonator array using an Ikeda-map approach, which evolves the intracavity fields round trip by round trip and therefore does not invoke the mean-field approximation~\cite{Hansson2015, Hashemi2024, Hashemi2025}. We use normalized coordinates $z,\tau\in[0,1]$, with $\OR = 2\pi$, $v_g = 1$, and the field amplitude rescaled such that the Kerr nonlinear coefficient $\gamma = 1$. During propagation within each resonator, the field evolves according to
\begin{equation}
\frac{d E^{m}_{x,y}(z,\mu)}{dz} = i\!\left(\wp - \wo - \frac{D_2}{2}\mu^{2}\right) E^{m}_{x,y}(z,\mu)
- \frac{\alpha}{2}\, E^{m}_{x,y}(z,\mu)
+ i\!\int_{0}^{1}\!\! d\tau\,\lvert E^{m}_{x,y}(z,\tau)\rvert^{2}\, E^{m}_{x,y}(z,\tau)\, e^{i\w_{\mu}\tau},
\label{eq:ikeda_prop}
\end{equation}
where $E^{m}_{x,y}(z,\tau)$ denotes the field in the ring resonator at the lattice site $(x,y)$ during round trip $m$. The longitudinal modes are indexed by $\mu$, with resonance frequencies $\w_{0,\mu} = \wo + \OR\mu + (D_2/2)\mu^{2}$, where $\OR = D_{1} = 2\pi v_g/L_R$, $L_{R}$ is the resonator length and $D_2$ characterizes the second-order dispersion. $\alpha$ is the propagation loss. Equation~\eqref{eq:ikeda_prop} is integrated using a split-step method in a reference frame moving with the group velocity $v_g$.
 
At each coupling point $z_c$, the fields in two coupled rings are mixed by a $2\times2$ beam-splitter transformation
\begin{equation}
\begin{pmatrix} E^{m}_{x,y}(z_c^+,\tau) \\[3pt] E^{m}_{x',y'}(z_c^+,\tau) \end{pmatrix} =
\begin{pmatrix} t&i\kappa \\[3pt] i\kappa & t \end{pmatrix}
\begin{pmatrix} E^{m}_{x,y}(z_c^-,\tau) \\[3pt] E^{m}_{x',y'}(z_c^-,\tau) \end{pmatrix},
\label{eq:ikeda_coupling}
\end{equation}
where $z_c^-$ and $z_c^+$ denote the fields immediately before and after the coupler, respectively. The transmission coefficient $t = \sqrt{1-\kappa^{2}}$ and the coupling coefficient $\kappa = \sqrt{2J}$ is the same for the nearest and the next-nearest-neighbor couplings. 

The input-output (IO) ring, at $(x,y) = (1,1)$, is coupled to the pump/output waveguide as
\begin{equation}
E^{m}_{\mathrm{out}}(\tau) = i\kappa_{IO}\, E^{m}_{1,1}(z=1,\tau) + t_{IO}\, E_{\mathrm{in}}(\tau),
\label{eq:ikeda_io}
\end{equation}
where $E_{\mathrm{in}}$ is the continuous-wave pump ($\mu = 0$) and $E^{m}_{\mathrm{out}}$ the output
on round trip $m$. Iterating Eqs.~\eqref{eq:ikeda_prop}--\eqref{eq:ikeda_io} for a fixed pump frequency and pump power yields the dark-soliton, platicon, bright-like-soliton, or chaotic states.
 
For our choice of normalized device parameters, the physical pump power required to generate the comb is related to the normalized field $E_{\mathrm{in}}$ as
\begin{equation}
P_{\mathrm{in}} = \frac{c}{n_2\,\omega_0\,L_R}\,A_{\mathrm{eff}}\,\lvert E_{\mathrm{in}}\rvert^{2}.
\label{eq:pin}
\end{equation}

For representative Si$_3$N$_4$ resonator parameters, $\lambda_0 =  1550 ~\text{nm}$ ($\omega_0 = 2\pi c/\lambda_0$), $n_2 = 2.4\times 10^{-19} ~\text{m}^2/\text{W}$, group index $n_g = 2.1$ \cite{Luke2015}, an FSR of  250 GHz gives a ring length $L_R = c/(n_g\,\mathrm{FSR}) \simeq ~600 \mu \text{m}$. With effective mode area $A_{\mathrm{eff}} \simeq 1 ~\mu \text{m}^2$, the prefactor in Eq.~\eqref{eq:pin} is $\simeq 1.71\times 10^{3} ~\text{W}$.

For the dark solitons and platicons states, which we observe at $E_{\mathrm{in}}=0.0469$, the required pump power is then $P_{\mathrm{in}}\simeq 3.76 ~\text{W}$. For the bright-like solitons, which we observe at $E_{\mathrm{in}}=0.062$, required $P_{\mathrm{in}}\simeq 6.58 ~\text{W}$. These values are comparable to pump powers accessible in integrated Kerr-comb experiments and can be further reduced through lower propagation loss and optimization of the inter-resonator and input–output coupling strengths, $J$ and $\kappa_{IO}$, respectively. 



\providecommand{\noopsort}[1]{}\providecommand{\singleletter}[1]{#1}